# Response of Glass and Liquid Phases in the Vortex lattice to an external AC magnetic field at different frequencies

Massimiliano Polichetti, Maria Giuseppina Adesso, Sandro Pace

*Dipartimento di Fisica, Universita' degli Studi di Salerno*

*INFM/SUPERMAT Via S. Allende, 84081,*

Baronissi (SALERNO) ITALY

**Abstract**

We individuated a method to distinguish a glass phase from a highly viscous liquid phase in a lattice of vortices, established in type-two superconductors. Our analysis is based on the study of the temperature dependence of numerically obtained $1^{st}$ and $3^{rd}$ harmonics curves of the AC magnetic susceptibility, by changing the frequency of the applied AC magnetic field. The harmonics are obtained by integrating the non-linear diffusion equation for the magnetic field, with different voltage-current characteristics, corresponding to the two different phases. This method could be applied to the analysis of experimental curves in order to determine if the detected magnetic response of superconducting samples may be interpreted in terms of vortex glass or Kim-Anderson model.

**1. Introduzione**

Glassy systems are complex systems characterised by "constitutive objects" interacting between them, put in a no-clean environment, with quenched structural disorder [1,2]. The interactions of disorder with the constitutive objects compete with those of these objects with each other and no long-range order is present: this situation corresponds to a high-temperature "liquid" phase. Nevertheless, a "freezing" transition without an apparent symmetry breaking, can occur from the liquid to a low-temperature phase, similar to the "solid phase". This phase does not show a crystallization, but a new kind of order is present, in which the objects are randomly aligned: this is the "glass" phase. Examples of the glassy systems are in all the fields of the physics as well as in the common life [1]. One of the most studied is the "spin glass" [2]: a magnetic system in which the constitutive objects are the magnetic moments of the spins, that interact with each other and with the frozen-in structural disorder; this glass phase is characterized by the spins aligned in random directions. A way to evidence the presence of a spin glass system is to analyse the response of the sample to an AC magnetic field, as a function of the temperature [3], at different frequencies.

**2. The vortex lattice as a glassy system**

Another particular system in which we can evidence a glass phase is the vortex lattice in the mixed state of type-II superconductors [1]. In fact, the vortices interact with each other in the lattice and, if the sample is "clean" and no external current is added, the equilibrium is reached when the vortices are in an hexagonal lattice (Abrikosov lattice). Nevertheless, if there is an external flowing current (still in the clean limit) the arrangement of the vortices in the lattice keeps fixed and the long-range order is preserved, but still the vortices can move collectively and this motion causes a no-zero resistivity (called "Flux Flow resistivity", $\rho_{Flow}$). Moreover, also the thermal energy of the surrounding thermal bath causes the vortices to move, which gives rise to an associated non-linear resistivity ("Flux Creep resistivity" $\rho_{Creep}$). Finally, if the quenched disorder is present ("no-clean" limit), any impurities or defects interact with the vortices, both acting as a restraint to their movement and destroying the long-range order. The described situation corresponds to a complex system in the "liquid phase": here the "constitutive objects" are the vortices, whereas the pinning centers and the thermal fluctuations represent the disorder. The dynamic response of this kind of system is highly non-linear, with a voltage-current (V-I) characteristic exhibiting a critical current density, $J_c$, below which dissipation is strongly reduced. There are two opposite theories describing what happens for a current density $J < J_c$: the Kim-Anderson scenario [4], in which the vortices are still in a liquid phase (although highly viscous), and the dissipation is finite; and the Fisher's theory [5], with a resistivity approaching zero, that corresponds to the vortex glass phase. Some experimental evidences of the vortex glass phase have been obtained, by performing direct measurements of the V-I characteristics [6]. Nevertheless, there are some controversies about the interpretation of this kind of experimental data [1]. By analogy with the study of the spin glass systems, we will introduce a method in order to individuate the vortex glass phase, starting from the response of the vortices to an AC magnetic field, at different frequencies ν.

## 3. The numerical computation method

We simulated the 1$^{st}$ and the 3$^{rd}$ harmonics of the AC susceptibility, by integrating the non-linear equation describing the diffusion of the magnetic field B inside the sample:

$$\frac{\partial B}{\partial t} = \frac{\partial}{\partial x}\left[\left(\frac{\rho(B,J)}{\mu_0}\right) \cdot \frac{\partial B}{\partial x}\right], \qquad (1)$$

where $\mu_0 = 4\pi \cdot 10^{-7} H/m$, with the following dependence for the resistivity:

$$\rho(B,J) = \rho_{Flow}(B) \cdot e^{-\left(\frac{U_p(J,T)}{k_B T}\right)}. \qquad (2)$$

By changing the relation between the pinning potential and the current, $U_p(J)$, it is possible to alternatively analyse either the glass or the liquid phase. In particular, the Vortex Glass theory predicts that the pinning energy $U_p$ has a non-linear dependence on the current density, through the μ parameter [7]:

$$U_p(J) = \frac{U_0}{\mu}\left[\left(\frac{J_C}{J}\right)^\mu - 1\right], \qquad (3)$$

where $U_0$ is the pinning potential height at zero current. In the framework of the Collective-Creep theory [1], μ has different values depending on the vortices dynamical arrangement: 1/7 for systems in which the vortices participate to the magnetic diffusion as isolated vortices ("Single Vortex" regime); 3/2 for vortices grouped in "small bundles", and 7/9 for "large bundles". On the contrary, the Eq. 3 with μ=-1 corresponds to the liquid phase described in the linear Kim-Anderson model [7].

## 4. Linear and non-linear response to the AC magnetic field in the Vortex glass phase and in the Kim-Anderson model

In Fig. 1 the temperature dependence of the 1$^{st}$ harmonics, at different frequencies of the AC magnetic field are shown, as simulated in the Single Vortex regime, at a fixed amplitude of the AC magnetic field, $h_{AC}$, and without a DC magnetic field ($H_{DC}$=0). Similar results have been obtained for Small Bundle and Large Bundle regimes (not reported). The same conditions are used to simulate the 1$^{st}$ harmonics in the framework of the Kim-Anderson

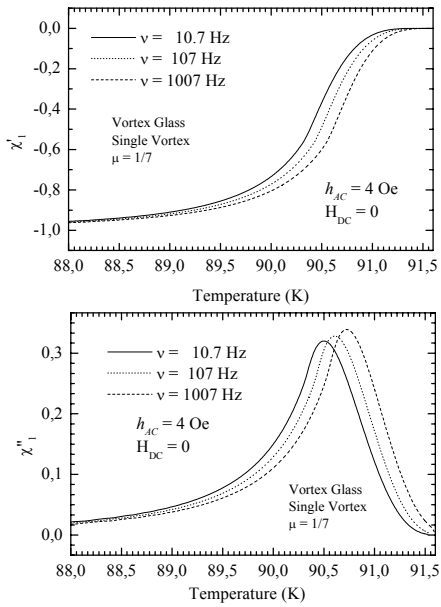

Fig.1 Temperature dependence of 1$^{st}$ harmonics at different frequencies, simulated by using the Vortex glass model in the Single Vortex regime.

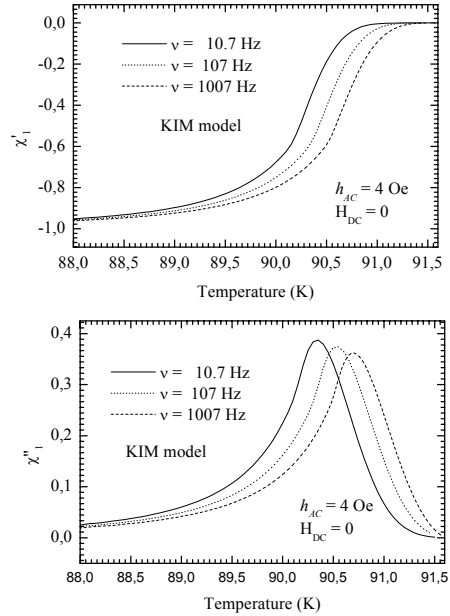

Fig.2 Temperature dependence of 1$^{st}$ harmonics at different frequencies, simulated by using the Kim-Anderson model.

model, as reported in Fig.2.

From Fig.1, we can observe that, in the glass phase if the frequency is increased, the temperature of the peak in the imaginary part of the 1$^{st}$ harmonics, $T_p$, shifts towards higher temperature and the height of the peak, $\chi_1^{"}(T_p)$, grows. Nevertheless, in the Kim-Anderson model (Fig.2), if $\nu$ is increased, $T_p$ also shifts towards higher temperature, but $\chi_1^{"}(T_p)$ decreases. Therefore, by comparing the results obtained in the glass phase with those in the "liquid phase", it is possible to distinguish between the two phases simply by analysing the frequency behaviour of the $\chi_1^{"}(T_p)$. This is evidenced also in Fig. 3, where the $\chi_1^{"}(T_p)$ as a function of the frequency, simulated for $1 \leq \nu \leq 10000$ in all the considered models, is shown.

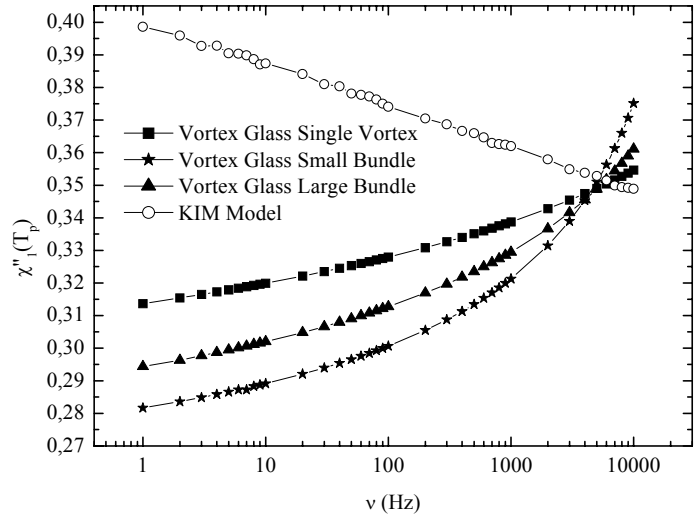

Fig.3 Heights of the peaks in the imaginary part of the 1st harmonics as a function of the frequency, simulated with the vortex glass model (in all the regimes), and in the Kim-Anderson scenario.

In particular, in Fig.3 we can notice that $\chi_1''(T_p)$ has a logarithmic decreasing frequency behaviour in the Kim model, which is completely different from the increasing, highly non-linear dependence in the glass phase.
Nevertheless, the qualitative behaviour of the real part of the AC magnetic susceptibility is similar in both the glass and the liquid phase.
A similar analysis can be faced by considering the higher harmonics of the AC magnetic susceptibility. In particular, in Fig.4, the 3rd harmonics of the AC susceptibility simulated in the glass phase (in the single vortex regime), are shown, for different frequencies. In Fig.5 the corresponding curves obtained by using the Kim-Anderson model are reported.

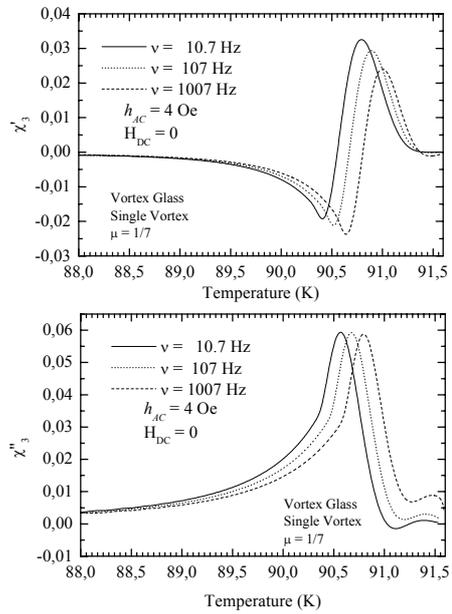

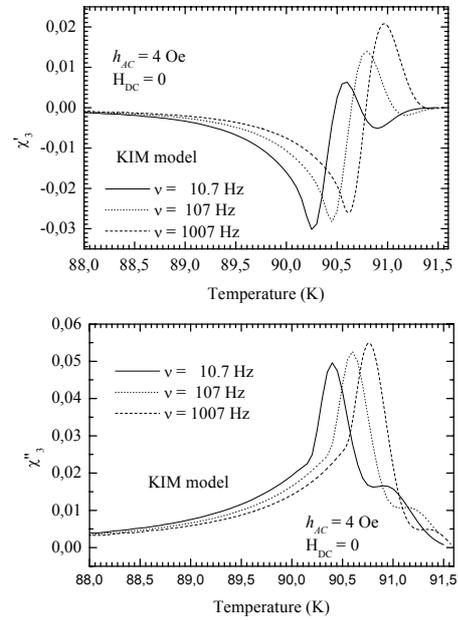

Fig.4 Temperature dependence of $3^{rd}$ harmonics at different frequencies, simulated by using the Vortex glass model in the Single Vortex regime.

Fig.5 Temperature dependence of $\chi_3$ at different frequencies, simulated by using the Kim-Anderson model.

The principal differences between the glass and the liquid phase can be evidenced if we observe the real part, $\chi_3'(T)$, of the 3rd harmonics. In particular, in all the considered models $\chi_3'(T)$ shows a minimum and a maximum, both depending on frequency. Nevertheless, for increasing frequencies, the glass phase shows that the absolute value of the minimum grows and the maximum decreases, whereas in the liquid phase their behaviour is opposite.

Finally, from the combined analysis of the 1st and 3rd harmonics curves [8], and as confirmed by the curves in Fig. 3, it is possible to affirm that, for increasing frequencies, the glassy system evolves from a Critical State Regime [9], in which the dissipations, proportional to $\chi_1''(T)$, are only hysteretic and minimal, towards a situation in which the flux dynamics effects become relevant in the magnetic response, and therefore the dissipative phenomena grow up.

On the contrary, the liquid phase appears as characterised by strong dynamic effects at low frequencies [8], which diminish as the frequency grows. This can be explained if we consider that, when the frequency is increased, the time window in which the dynamic phenomena (relevant in the liquid phase) can be detected is restricted.

## 5. Conclusions

We compared the vortex glass phase with a highly viscous liquid phase of the vortex lattice, in a type-two superconductor, starting from the analysis of the linear and non linear response of the sample to an applied AC magnetic field. In particular, we simulated the 1st and the 3rd harmonics of the AC magnetic susceptibility, at different frequencies, by choosing the vortex glass and the Kim-Anderson model. We observed that it is possible to distinguish the two phases by a simple measurement of the imaginary part of the 1st harmonics, because the height of the peak as a function of the frequency has a completely different behaviour into the two phases. Analogously, an opposite behaviour of the minimum and the maximum, present in the real part of the 3rd harmonics, can also be used as a good method in order to differentiate the two phases in the system. Finally, the effects on the dynamics, due to the application of an AC magnetic field on the sample, have been evidenced for the different phases. Therefore, the individuated differences between the two reported dynamic phases suggest that this kind of analysis could be a valid method for the interpretation of the experimental 1st and 3rd harmonics curves, like for example those reported in the references [10].

**References**

[1]  G. Blatter et al., Rev. Mod.Phys., 66 (1994) 1125.
[2]  K. Binder et A.P. Young, Rev. Mod.Phys., 58 (1986) 801.
[3]  Mulder et al., Phys. Rev. B, 23 (1981) 1384.
[4]  P.W. Anderson, Phys. Rev. Lett., 9 (1962) 309.; P.W. Anderson and T.B. Kim, Rev. Mod. Phys., 36 (1964) 89.
[5]  M.P.A. Fisher, Phys. Rev. Lett., 62 (1989) 1415.
[6]  R. H. Koch et al., Phys. Rev. Lett., 63 (1989) 2586.
[7]  H. H. Wen et al., Phys. Rev. Lett., 8 (1997) 1559.
[8]  M. Polichetti et al., Eur. Phys. J. B, 36 (2003) 26.
[9]  C.P. Bean, Phys. Rev. Lett., 8 (1962) 250.
[10] X. C. Jin et al., Phys. Rev. B, 47, 6982 (1993)